# GHz-gated InGaAs/InP single-photon detector with detection efficiency exceeding 55% at 1550 nm

L. C. Comandar[1,2], B. Fröhlich[1,a)], J. F. Dynes[1], A. W. Sharpe[1], M. Lucamarini[1], Z. L. Yuan[1], R. V. Penty[2] and A. J. Shields[1]

[1]Toshiba Research Europe Ltd, 208 Cambridge Science Park, Milton Road, Cambridge CB4 0GZ, United Kingdom

[2]Engineering Department, Cambridge University, 9 J J Thomson Ave, Cambridge CB3 0FA, United Kingdom

We report on a gated single-photon detector based on InGaAs/InP avalanche photodiodes (APDs) with a single-photon detection efficiency exceeding 55% at 1550 nm. Our detector is gated at 1 GHz and employs the self-differencing technique for gate transient suppression. It can operate nearly dead time free, except for the one clock cycle dead time intrinsic to self-differencing, and we demonstrate a count rate of 500 Mcps. We present a careful analysis of the optimal driving conditions of the APD measured with a dead time free detector characterization setup. It is found that a shortened gate width of 360 ps together with an increased driving signal amplitude and operation at higher temperatures leads to improved performance of the detector. We achieve an afterpulse probability of 7% at 50% detection efficiency with dead time free measurement and a record efficiency for InGaAs/InP APDs of 55% at an afterpulse probability of only 10.2% with a moderate dead time of 10 ns.

[a)] Electronic mail: bernd.frohlich@crl.toshiba.co.uk

I. INTRODUCTION

Single-photon detectors at telecom wavelengths have been the subject of intense research due to their application in a number of areas such as quantum key distribution (QKD) [1, 2], light detection and ranging (LIDAR) [3], or optical time-domain reflectometry (OTDR) [4]. Ideally, a single-photon detector should have a high detection efficiency with no significant dead time, low detection noise, and precise time resolution [5, 6, 7]. At the same time it should be compact with minimal requirement on cooling. Superconducting nanowire detectors satisfy most of these characteristics, however, they have the disadvantage of requiring cryogenic cooling [8, 9, 10]. InGaAs/InP avalanche photodiodes (APDs) instead need to be cooled only moderately and can even be used at room temperature [11]. However, their maximum detection efficiency is typically limited by high dark count probabilities and strong afterpulse noise [12].

Detectors based on APDs can be operated either in free-running or gated mode. The former can be used for applications where the arrival time of the photons is not precisely known and therefore is general purpose. However, long hold-off times have to be used to mitigate afterpulse effects which unavoidably reduce the maximum count rate [13]. For synchronous applications an APD is often operated in gated mode to allow rapid quenching of the avalanche current so as to reduce the avalanche charge flow and hence the detector recovery time. Gated mode operation is further improved with novel techniques such as sine-wave [14] or self-differencing (SD) filtering [15], or coherent addition of discrete harmonics [16]. All these techniques remove the capacitive response, arising from periodic gating, which would otherwise prevent the detection of weak avalanche signals. The SD method has the advantage of being independent of the detector gate waveform and driving conditions. Exploiting the periodic nature of the gating signal, a SD circuit subtracts the detector signal from two consecutive gates, which removes the capacitive response. The insensitivity towards changes of the detector gating signal facilitates optimization of the detector performance.

Here, we demonstrate that InGaAs/InP APDs operated in an optimized self-differencing mode can achieve detection efficiencies exceeding 55% – an unprecedented value for 1550 nm telecom wavelength APDs. Previous reports of efficiencies close to 50% have been measured at 1310 nm only [16, 17], where the quantum efficiency is higher than at 1550 nm by 25%. We use a dead time free detector characterization setup which allows precise measurement of performance parameters, such as detector afterpulsing. The highest efficiencies are obtained by reducing the gate width in combination with higher gate amplitudes, and importantly also elevated device temperature. A measurement of the external quantum efficiency (EQE) suggests that over 80% of the absorbed photons are detected under these operating conditions.

II. DEVICE PROPERTIES AND EXPERIMENTAL SETUP

The single-photon detection efficiency of an APD is determined by the product of four factors: the carrier absorption efficiency ($P_{ab}$) and transit probability ($P_{trans}$), dependent on the absorption material and device architecture, avalanche triggering probability ($P_{avl}$) determined by the bias conditions and device architecture, and the avalanche detection probability ($P_{det}$) determined by the external electrical circuitry [13]:

$$SPDE = P_{ab} \times P_{trans} \times P_{avl} \times P_{det}. \tag{1}$$

The product of the first two factors can be estimated from the I-V measurement shown in Fig. 1(a). At the punch-through (or reach-through) voltage, occurring around 36 V, the APD linear multiplication gain is close to unity. With unity gain, the external quantum efficiency (EQE) is determined by:

$$EQE = P_{ab} \times P_{trans} = \frac{I_{ph}}{P_{opt}} \frac{hc}{e\lambda}, \qquad (2)$$

where $I_{ph}$ represents the measured photocurrent, $P_{opt}$ is the incident optical power, $h$ is Planck's constant, $c$ the speed of light, $e$ the elementary charge and $\lambda$ the wavelength of the incident photons. It is evident from Fig. 1(a) that the gain has not completely reached a plateau at the punch-through voltage - which would indicate unity gain - for all device temperatures shown in the graph. Therefore, we upper bound the EQE to be approximately 69% at a device temperature of 20 °C representing the maximum achievable SPDE. The temperature dependency of the breakdown voltage ($V_{Br}$) is shown in the inset of Fig. 1(a). It increases linearly at a rate of 0.1 V/°C from 60.1 V at −50 °C to 67.2 V at 20 °C due to increasing phonon scattering [18].

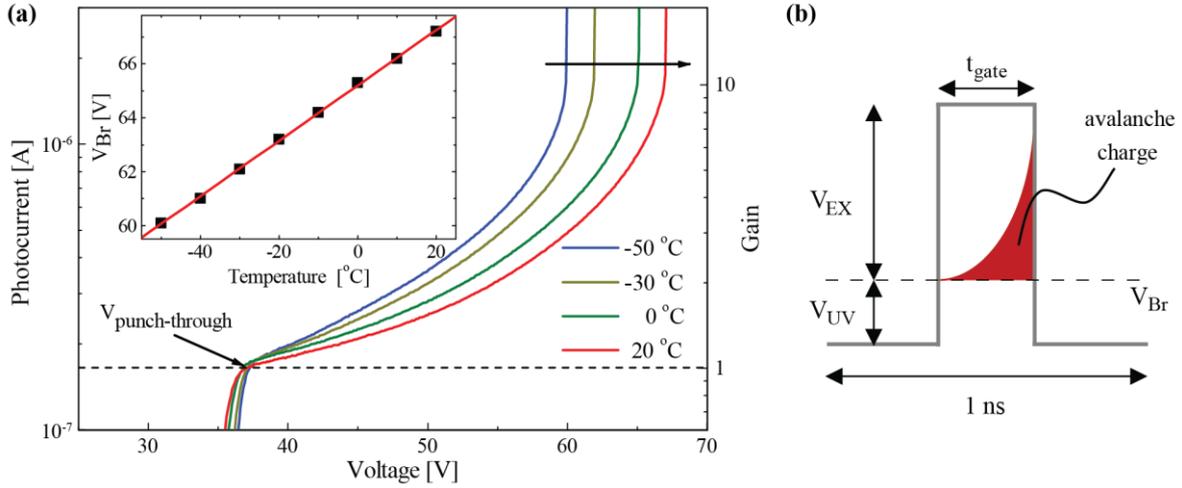

FIG. 1. a) Variation of the photocurrent with temperature. For these measurements the APD was illuminated with an optical power of 196 nW. Inset: Breakdown voltage dependency on temperature. b) Periodic biasing conditions of Geiger mode operation. Indicated in red is the evolution of current following the absorption of a photon and triggering of an avalanche. The current rises exponentially until it is quenched by reducing the voltage below breakdown. $V_{EX}$: Excess Voltage, $V_{UV}$: Under Voltage, $V_{Br}$: Breakdown Voltage, $t_{gate}$: gate duration.

APDs in Geiger mode are biased over the breakdown voltage for short periods of time ($t_{gate}$), as shown in Fig. 1(b), during which an initiated avalanche may grow into a detectable macroscopic current before being quenched by reducing the voltage below $V_{Br}$. Ideally, $t_{gate}$ should be optimized to limit the magnitude of the APD current and therefore avalanche charge. However, when the gate duration is in the nanosecond range the output of the diode is dominated by capacitive response with the avalanche buried within. Both of these signals depend on the APD operation and gate properties such as excess voltage ($V_{EX}$), under voltage ($V_{UV}$) and $t_{gate}$. Using the SD technique, which is largely insensitive to the waveform of the APD response [19], we are able to extract the avalanche signal and optimize the gate properties.

To accurately measure the performance of the device in gated mode we use a characterization method based on the technique described in Ref. [15]. A 1550 nm laser illuminates the APD with pulses 3 ps in duration at a repetition rate of 20 MHz. Pulses much shorter than the detector time response are required for a precise characterization of the detector. The pulses are attenuated to a photon level of μ = 0.1 photons/pulse with an uncertainty of ±5% given by the optical power meter and reproducibility of fiber connections. The output of the SD-APD gated at 1 GHz is discriminated by setting a threshold level and counts are recorded by dead time free electronics with a resolution of

100 ps and sorted into histograms. Measurements with and without illumination permit accurate determination of the dark count probability ($P_d$) and afterpulse probability ($P_a$). The SPDE η can be calculated using the following expression [15, 17]:

$$\eta = \frac{1}{\mu} \ln\left[\left(1 - \frac{R_d}{f_g}\right) \Big/ \left(1 - \frac{R}{f_l}\right)\right]. \tag{3}$$

where $R_d$ and $R$ are the dark count and detection count rates during the non-illuminated and illuminated gates, respectively, while $f_g$ and $f_l$ are the gating and laser repetition frequency.

III. RESULTS AND DISCUSSION

Using the characterization setup described above we determine the best driving conditions of the detector by varying the duration and peak-to-peak amplitude of the APD gating signal. Figure 2 summarises results of these measurements taken at a gating frequency of 1 GHz with zero acquisition dead time and with the detector cooled to −50 °C. To determine the optimal gate duration we vary the gate length and measure the afterpulse and dark count probability at several selected SPDEs (see Fig. 2(a)). For these measurements the AC driving signal is generated by an arbitrary waveform generator followed by a broadband electrical amplifier. The output amplitude of the arbitrary waveform generator is set constant and it produces an AC signal amplitude set to 7 V after amplification for a pulse duration of 500 ps. We only adjust the DC bias to achieve a given SPDE. Starting from long gate durations, $P_a$ and $P_d$ first decrease when the gate is shortened. Shorter gates lead to lower avalanche charge and therefore lower afterpulsing. At the same time the efficiency decreases which has to be compensated with higher excess voltage values, however, overall the detector performance improves. This is true for the dark count rate as well, which improves due to a decreasing time window for a thermally generated carrier to start an avalanche. For short active gate durations, especially 200ps, the pulse amplitude reduces due to the limited bandwidth of the driving electronics resulting in distorted and lower pulse amplitudes and therefore reduced SPDE. We found that a gate duration of approximately 360 ps as suggested by the data is optimal for our setup even for higher signal amplitudes and higher operating temperatures.

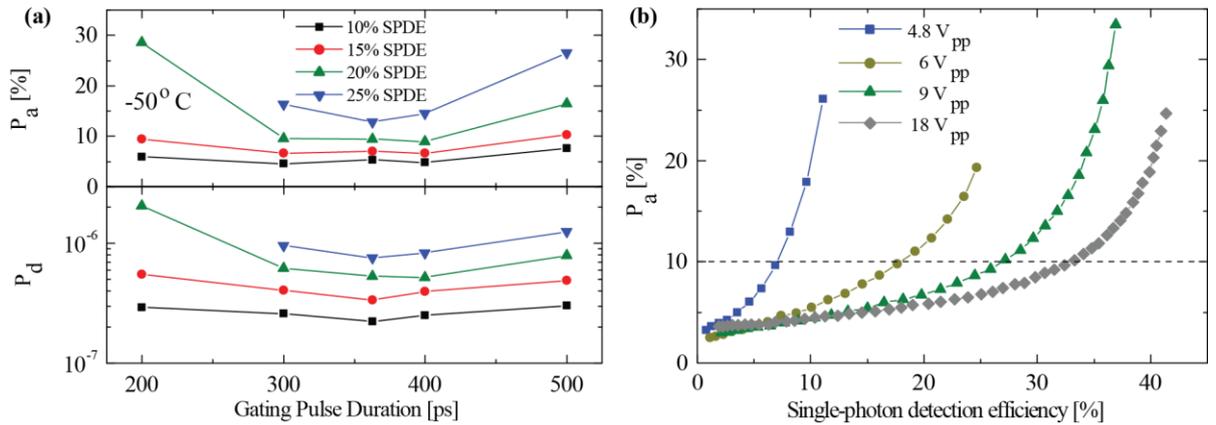

FIG. 2. a) Dark count probability (bottom) and afterpulse probability (top) as a function of the gating pulse duration for different SPDEs fixed by adjusting the applied DC bias. b) Afterpulse probability as a function of SPDE for different peak-to-peak voltages and fixed gate duration.

To optimise the peak-to-peak amplitude of the gating signal we keep the pulse duration constant at the optimum value (360 ps) and vary the amplitude between 4.8 V and 9 V using a 25 GHz bandwidth amplifier (see Fig. 2(b)). For each amplitude setting we measure the afterpulse probability as a function of the SPDE by changing the excess bias applied to the APD. The measurement shows that at a given efficiency afterpulsing decreases with higher gating pulse amplitude. This allows for SPDEs over 25% while keeping $P_a$ below 10% (black dashed line in Fig. 2(b). Overall, this data suggests that the SPDE could be further enhanced by increasing the amplitude to even higher voltages. To test this assumption, we use a higher power (6.5 W) but lower bandwidth amplifier (0.8 - 4.2 GHz) and increase the gating signal amplitude to 18 V while keeping the optimum gating pulse duration before the amplifier. In this configuration, at −50 °C, the detection efficiency improves to around 32% for the same afterpulse probability of less than 10% although the driving signal is distorted due to the reduced amplifier bandwidth.

Following the optimization of gate width and amplitude, we turn our attention to a third system parameter permitting further improvements to the detector performance, the operating temperature. In Fig. 3 we plot the dependence of afterpulse and dark count probability on the SPDE, for different operating temperatures from −50 °C to 20 °C. Again, the single photon detection efficiency is tuned by varying the DC bias. The data shows that for temperatures above −30 °C the dark count probability exhibits an approximate doubling every 10 °C. Below this temperature the dark count rate is limited by effects related to high APD gating amplitudes [20]. At the same time, the point at which the afterpulse probability starts to rise dramatically is shifted to higher efficiencies at higher temperatures. At a given efficiency, the afterpulse noise decreases with increasing temperature. Therefore, higher device temperatures do not necessarily lead to an increased noise level. For high count rate applications, operation at elevated temperatures can even be beneficial [11].

At a fixed afterpulse probability of 10% we observe that the SPDE increases linearly with temperature as shown in the inset, similarly to the linear increase of the breakdown voltage. Various effects leading to a change of efficiency with temperature such as the changing voltage gap between punch-through voltage and breakdown voltage [21] or an improvement of InGaAs absorption coefficients [22] have been described in literature. However, the strength of the effect (>50% increase from −50 °C to 20 °C) is surprising and the exact cause requires further investigation. Due to our dead time free characterization method we do expect the afterpulse probability to be influenced by the carrier de-trapping time only on a very short timescale of a few nanoseconds. Here, the intrinsic 1 ns dead time of the self-differencing method can decrease the afterpulse probability when the decay time reduces. However, this effect does not account for the strong decrease in afterpulsing we measure. Instead, a reduction in the carrier trapping probability following an avalanche could account for the decrease in afterpulsing. The SPDE rises to over 55% at room temperature which is the highest reported with an APD at a wavelength of 1550 nm to date. The SPDE also reaches 80% of the EQE measured to be 69% at unity gain. We achieve this performance with an excess bias of approximately 9.5V. The main uncertainty to the efficiency value comes from the calibration of the laser power, which is exact to ±5%. However, we expect the power to be overestimated in general (and therefore the SPDE to be underestimated) due to the final fiber connection between laser and APD.

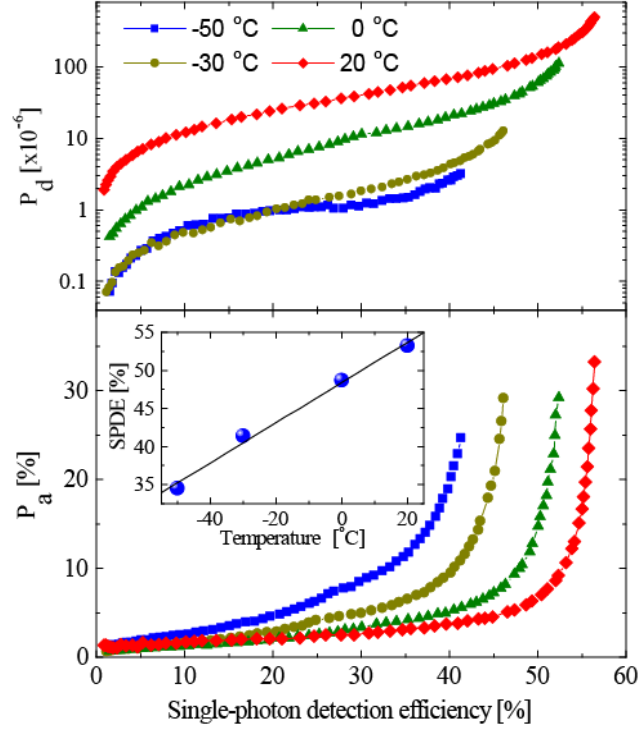

FIG. 3: Dark count probability and afterpulse probability as a function of the SPDE for temperatures in the range of −50 °C to 20 °C at an applied gate amplitude of 18 V. Inset: The variation of the SPDE at an afterpulse probability of 10% with temperature.

In addition to system parameters such as driving conditions and temperature, post-processing by applying a dead time after each detection event can also be used for detector optimization. In Fig. 4(a) we compare the afterpulse probabilities for various SPDEs at room temperature, with and without applying a software dead time after each discriminated avalanche. At an SPDE of over 50% the afterpulse probability drops to 4.2% as compared to 7% in the dead time free case. The inset of Fig. 4(a) shows the histogram of the illuminated and non illuminated gates at an SPDE of 55% with and without dead time highlighting the contribution made by dark counts and the reduction of afterpulsing. In the histogram without post-processing dead time applied the influence of the self-differencing method can be seen. It leads to a reduction of counts in the first gate after the illuminated gate, which can be understood as an intrinsic 1 ns dead time. We note that the use of a dead time will limit the maximum count rate. However, for our chosen dead time of 10 ns the reduction of the detector count rate is less than 10% up to a count rate of 10 Mcps.

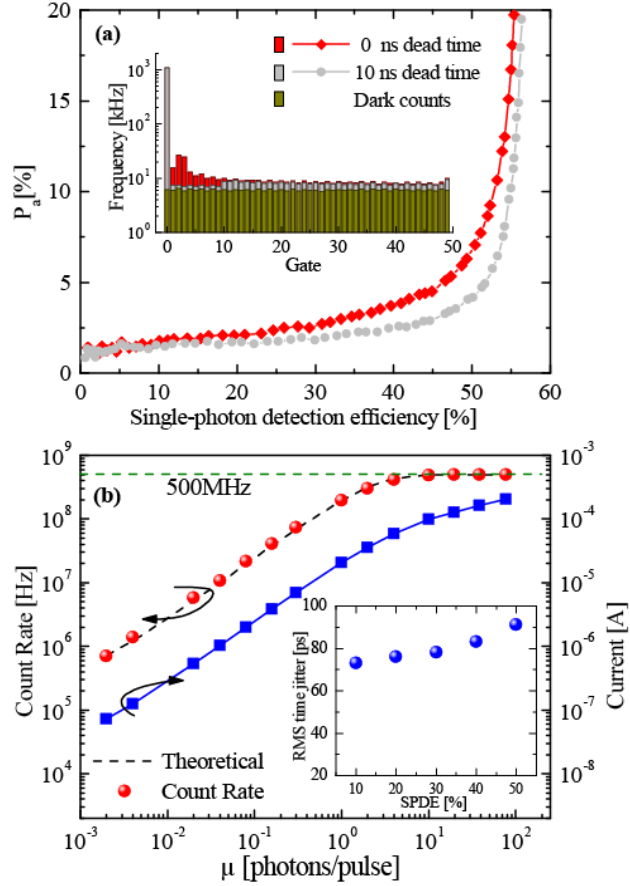

FIG. 4. a) Comparison of $P_a$ values with and without applying a 10ns dead time after each count at a device temperature of 20 °C. Inset: Histograms acquired over one second divided into 1 ns bins recorded without dead time (red) and with 10 ns dead time (grey) at an SPDE of 55%. Also shown is a histogram without illumination (dark yellow). b) Measured count rate (red circles) and APD current (blue squares) for increasing photon flux at a gating frequency of 1 GHz and a laser repetition rate of 500 MHz. The dashed line represents the expected photon count rate following Eq. 4 and the blue line is a guide to the eye. Inset: The RMS time jitter at different SPDEs.

In applications where a high count rate is essential, no dead time should be applied. Previously, SD detectors have reached 1 Gcps at an efficiency of 20% [23] and a gating frequency of 2 GHz. Ideally the detector should maintain its high count rate capability even at the high efficiencies demonstrated here. From Eq. (3), we can estimate the detection rate as:

$$R = f_l \left[1 - e^{-\mu\eta}(1 - P_d)\right]. \qquad (4)$$

We illuminate the detector with a laser pulsed at 500 MHz, and the APD is biased to have an SPDE of 50% in the low flux regime. Fig 4(b) shows the linear increase of the count rate with the photon flux before saturation. This indicates that the SPDE of 50% is independent of the count rate making the SD detector desirable for high count rate applications. Due to the consecutive gate cancellation approach employed by the SD circuitry, the maximum count rate is half of the gating frequency. We have reached this theoretical maximum count rate even at high detection efficiency suggesting no deterioration of the detection efficiency in the high photon flux regime. This is further evidenced by a measurement of the APD current shown in the same graph. It exhibits the same linear trend as the count rate. The avalanche charge, which is the ratio of current divided by count rate, therefore stays constant until shortly before the saturation point.

To characterize the detector time resolution we measured the root-mean-square (RMS) value of the photon times of arrival, defining the detector time jitter, with a 8 ps resolution time-to-digital converter. We chose to plot the RMS value due to the non-Gaussian shape of the distribution of photon arrival times which changes with efficiency. The inset in Fig. 4(b) shows how the jitter changes with SPDE with the detector operated under optimum conditions at 20 °C. It increases only slightly from 73 ps to 91 ps from 10% to 50% SPDE demonstrating that the detector maintains a good time resolution also at the highest efficiencies.

IV. SUMMARY

In conclusion, we have demonstrated a high efficiency single-photon detectors based on InGaAs/InP APDs for photon detection at the wavelength of 1550 nm. Most notably our investigation shows that the maximum achievable SPDE at a given afterpulse probability increases linearly with temperature. Utilizing this effect and using optimized biasing conditions we achieve a record SPDE for APD-based detectors of over 55% at 1550 nm. A measurement of the EQE suggests that over 80% of the absorbed photons are detected, showing the efficiency of the technique in approaching the performance limit of the APD. Our detector operates dead time free except for the one clock cycle dead time intrinsic to the SD technique and we have demonstrated a count rate of 500 MHz making the SD-APD suitable for a number of high speed applications such as QKD, LIDAR, or OTDR.

**Acknowledgements**

L. C. Comandar acknowledges personal support via the EPSRC funded CDT in Photonics System Development.